\documentclass[12pt,preprint]{aastex}

\shorttitle{Bars, Nuclear Spirals and Fueling AGN}
\shortauthors{Martini, Regan, Mulchaey, \& Pogge}

\slugcomment{ApJ Accepted [Dec 17 2002]}

\begin{document}

\newcommand{\um}{\mu{\rm m}}
\newcommand{\hst}{{\it HST\,\,}}
\newcommand{\eg}{{\rm e.g.\,}}
\newcommand{\kms}{{\rm \,km\,s^{-1}}}
\newcommand{\etal}{{\rm et al.\,}}

\title{Circumnuclear Dust in Nearby Active and Inactive Galaxies. II.  Bars, 
Nuclear Spirals, and the Fueling of Active Galactic Nuclei\altaffilmark{1}}

\altaffiltext{1}{Based on observations with the
NASA/ESA {\it Hubble Space Telescope} obtained at the the Space Telescope
Science Institute, which is operated by the Association of Universities for
Research in Astronomy, Incorporated, under NASA contract NAS5-26555.}

\author{Paul Martini \altaffilmark{2}, 
Michael W. Regan \altaffilmark{3} 
John S. Mulchaey \altaffilmark{2}, 
Richard W. Pogge\altaffilmark{4}, 
}

\altaffiltext{2}{Carnegie Observatories, 813 Santa Barbara St.,
Pasadena, CA 91101-1292, martini@ociw.edu, mulchaey@ociw.edu}

\altaffiltext{3}{Space Telescope Science Institute, 3700 San Martin Drive, 
Baltimore, MD 21218, mregan@stsci.edu}

\altaffiltext{4}{Department of Astronomy, Ohio State University,
140 W. 18th Ave., Columbus, OH 43210, pogge@astronomy.ohio-state.edu}

\begin{abstract}

We present a detailed study of the relation between circumnuclear dust 
morphology, host galaxy properties, and nuclear activity in nearby galaxies.
We use our sample of 123 nearby galaxies with visible--near-infrared 
colormaps from the {\it Hubble Space Telescope} to create well-matched,  
``paired'' samples of 28 active and 28 inactive galaxies, as well as 19 barred 
and 19 unbarred galaxies, that have the same host galaxy properties. 
Comparison of the barred and unbarred galaxies shows that grand design 
nuclear dust spirals are only found in galaxies with a large-scale bar. These 
nuclear dust spirals, which are present in approximately a third of all barred 
galaxies, also appear to be connected to the dust lanes along the leading edges 
of the large-scale bars. Grand design nuclear spirals are more common 
than inner rings, which are 
present in only a small minority of the barred galaxies. 
Tightly wound nuclear dust spirals, in contrast, show a strong tendency to 
avoid galaxies with large-scale bars. 
Comparison of the AGN and inactive samples shows that nuclear dust spirals, 
which may trace shocks and angular momentum dissipation in the ISM, occur with 
comparable frequency in both active and inactive galaxies. 
The only difference between the active and inactive galaxies is that several 
inactive galaxies appear to completely lack dust structure in their 
circumnuclear region, while none of the AGN lack this structure. 
The comparable frequency of nuclear spirals in active and inactive galaxies, 
combined with previous work that finds no significant differences 
in the frequency of bars or interactions between well-matched active and 
inactive galaxies, suggests that no {\it universal} fueling mechanism for 
low-luminosity AGN operates at spatial scales greater than $\sim 100$ pc 
radius from the galactic nuclei. The similarities of the circumnuclear 
environments of active and inactive galaxies suggests that the lifetime of 
nuclear activity is less than the characteristic inflow time from these 
spatial scales. An order-of-magnitude estimate of this inflow time is the 
dynamical timescale. This sets an upper limit of several million years to the 
lifetime of an individual episode of nuclear activity.  

\end{abstract}


\section{Introduction}

Many observational programs over the past few years have led to the proposition 
that all galaxies with a substantial spheroid component contain supermassive 
black holes, irrespective of the presence or absence of nuclear activity 
\citep[\eg][]{richstone98,ferrarese00,gebhardt00}. 
Since black holes grow via the accretion of matter and this accretion leads 
to detectable nuclear activity, these results imply that all galaxies must 
go through an accretion phase, yet the mechanism which triggers nuclear 
activity in ``currently'' active galaxies remains unknown. 
In order to fuel active galactic nuclei (AGN), essentially all of the angular 
momentum must be removed from some fraction of the host galaxy's interstellar 
medium (ISM). 
Low-luminosity AGN, which dominate the local population, 
require accretion rates of $0.01 - 0.1 M_\odot$ yr$^{-1}$, assuming typical 
radiative efficiencies.

Studies of AGN and inactive control samples have investigated the frequency 
of several mechanisms for angular momentum transport to determine their 
viability. 
Interactions between galaxies is one good candidate \citep{toomre72} 
as theoretical simulations of mergers show significant accretion into the 
central regions of the merger remnant \citep[\eg][]{hernquist89,barnes91,
mihos96}.
Interactions may be responsible for triggering AGN activity
in the more luminous quasars \citep{hutchings97,bahcall97},
yet detailed studies of interacting pairs have not found a statistically
significant excess
of the lower-luminosity Seyfert galaxies in interacting systems
\citep{fuentes88}.
Large-scale bars have also been proposed as a mechanism to fuel nuclear 
activity \citep{simkin80,schwarz81}. 
The nonaxisymmetric potential due to a large-scale bar leads to the formation
of a shock front along the bar's leading edges 
\citep{prendergast83,athanassoula92} and material has been observed 
flowing into the central regions of several barred galaxies 
\citep{quillen95,regan97,jogee99}.
However, detailed near-infrared (NIR) studies of large samples of active and 
inactive galaxies have shown either no, or at most a marginal ($2.5\sigma$), 
excess of large-scale bars in active samples \citep{mulchaey97b,knapen00}.
These studies of interacting and barred galaxies pushed the effective spatial
resolution limit of ground-based observations for large samples of AGN, yet 
the typical spatial resolution of these investigations remain many hundreds of 
parsecs. 

Several \hst programs over the past few years have targeted the circumnuclear 
morphology of large active galaxy samples to search for signatures of AGN 
fueling \citep[\eg][]{malkan98,regan99,martini99}.  
One of the main goals of these programs was to investigate the fraction of 
Seyferts with nuclear bars (bars with semimajor axis lengths typically less 
than a kiloparsec), which could be comprised of gas or stars 
\citep{shlosman89,pfenniger90} and cause the transport of matter from 
approximately a kiloparsec to tens of parsecs. 
However, these studies have found nuclear bars in only $\sim 25$\% of all 
Seyferts \citep{regan99,martini01} and studies of Seyfert and control 
samples have found similar fractions of double bars in samples of active and 
inactive galaxies with large-scale bars \citep{marquez00,laine02,erwin02}. The 
comparable fractions of nuclear bars in active and inactive galaxies, combined 
with the apparent absence of them in the majority of all active galaxies, 
suggests that some other mechanism is needed to fuel nuclear 
activity in many active galaxies. 

One new candidate that arose from the \hst studies is nuclear dust spirals
\citep{regan99,quillen99,martini99,pogge02}. 
Visible--NIR color maps of 
the majority of the active galaxies in these surveys showed nuclear 
spirals, with a wide range of coherence, that extend from approximately a 
kiloparsec down to tens of parsecs (the limiting spatial resolution of the 
nearest subsample).
These nuclear spirals are distinct from the spiral arms in the main disks of
these galaxies as they appear to have density contrasts of only a factor of a
few above the ambient ISM and no associated star formation. 
Nuclear spirals are a promising fueling mechanism not only by virtue of their 
frequency, but also because they may mark the location of shock fronts 
or turbulence in the central, circumnuclear gaseous disks
and therefore trace the sites of angular momentum dissipation.  
The possibility of shock-driven inflow, as traced by nuclear spiral structure, 
has been the subject of a number of recent theoretical studies 
\citep{fukuda98,montenegro99,wada01a,maciejewski02,wada02}. 

While most of the observational programs to date have targeted the 
circumnuclear region of active galaxies, nuclear dust spirals have also been 
observed in a small number of inactive galaxies with single-bandpass 
observations \citep{phillips96,carollo98}. 
In \hst Cycle~9 we began a program 
(SN8597, PI Regan) to obtain WFPC2 images of galaxies with prior NICMOS 
observations (from SN7330, PI Mulchaey and GO7867, PI Pogge) in order to 
quantify the frequency of nuclear spiral structure in inactive galaxies. 
We present the observations of our final sample of 123 galaxies, along with a
description of the sample, survey design, and classification system for 
circumnuclear dust structure, in \citet[][hereafter Paper~I]{martini02}. 
Our nuclear dust classification has six types, including four for nuclear 
dust spirals: grand design, tightly wound, loosely wound, and chaotic spirals. 
We placed galaxies with dust structures but without evidence for nuclear 
spirals in a fifth, ``chaotic'' class, and 
galaxies with no detected circumnuclear dust structure into a sixth,  
``no structure'' class. 
The final dataset presented in Paper~I, in spite of the initial effort to 
create a well-match active and control sample, is relatively heterogeneous 
due to both the vagarious \hst snapshot scheduling and our attempt to 
augment the sample with additional nearby galaxies of interest. 

In the present paper we create well-matched subsamples of the full dataset 
presented in Paper~I in order to measure the relative frequency of nuclear 
dust spirals in active and inactive galaxies. 
This sample creation, described in the next section, draws unique pairs of 
active/inactive or barred/unbarred galaxies that otherwise have all of the 
same host galaxy properties and are at the same distance, while also maximizing 
the total size of the subsample drawn from the full dataset. In 
\S\S\ref{sec:active} and \ref{sec:all} we describe the trends for 
circumnuclear dust morphology for the active/inactive and 
barred/unbarred samples, respectively. The implications of these results 
for the fueling of nuclear activity are described in \S\ref{sec:discuss} and 
our results are summarized in \S\ref{sec:conc}. 

\section{Paired Sample Selection} \label{sec:match}

To construct the best-matched sample given the available observations, 
we have developed an algorithm 
to identify a unique inactive, control galaxy for each active galaxy and 
maximize the 
number of pairs of active and inactive galaxies. The host galaxy properties 
used to identify possible control galaxies for each active galaxy were the 
four properties used to define the original RSA control sample: 
Hubble type ($T$), blue luminosity ($M_B$), heliocentric velocity ($v$), and 
inclination ($R_{25}$), along with a fifth property: the angular size of the 
host. 
This fifth property is expressed as the fraction ($frac_{20}$) of the angular 
extent ($D_{25}$) of the host galaxy within the $20''$ 
field of view of the color maps shown in Figure~1 of Paper~I. 
The constraints on these parameters are: $\Delta T < 1$, 
$\Delta M_B < 0.5$ mag, $\Delta v < 1000 \kms$, $\Delta R_{25} < 0.1$, and 
$\Delta frac_{20} < 0.1$. 
These constraints were applied to the subsample of 86 (out of 123) galaxies 
with $v < 5000 \kms$ and $R_{25} < 0.30$. These two cuts were chosen to 
avoid distant galaxies, where the \hst spatial resolution is comparable to 
ground-based observations of nearby galaxies, and highly inclined systems, 
where we found our morphological classifications were less reliable 
(see Paper~I). 

The algorithm first identifies all possible control galaxies for each active 
galaxy through measurement of the absolute difference between each of the 
host galaxy parameters. 
As there are more active than inactive galaxies in the final dataset, 
the number of inactive galaxies is the main constraint on the final 
number of matched pairs. 
The first step in the algorithm is therefore to identify each inactive galaxy 
that has only one active galaxy match.  
These unique galaxy pairs are placed into the output catalog and the active 
galaxies in these pairs are then removed from the list of possible matches for 
the remaining inactive galaxies. 
This procedure is then repeated until no potential matched pairs remain. 
For the case where there are no inactive galaxies with a unique active galaxy 
match, the control galaxy with the fewest number of active galaxy matches 
was picked and matched with the active galaxy with the smallest number of 
other control galaxy matches. 

\begin{figure}
\epsscale{0.70}
\plotone{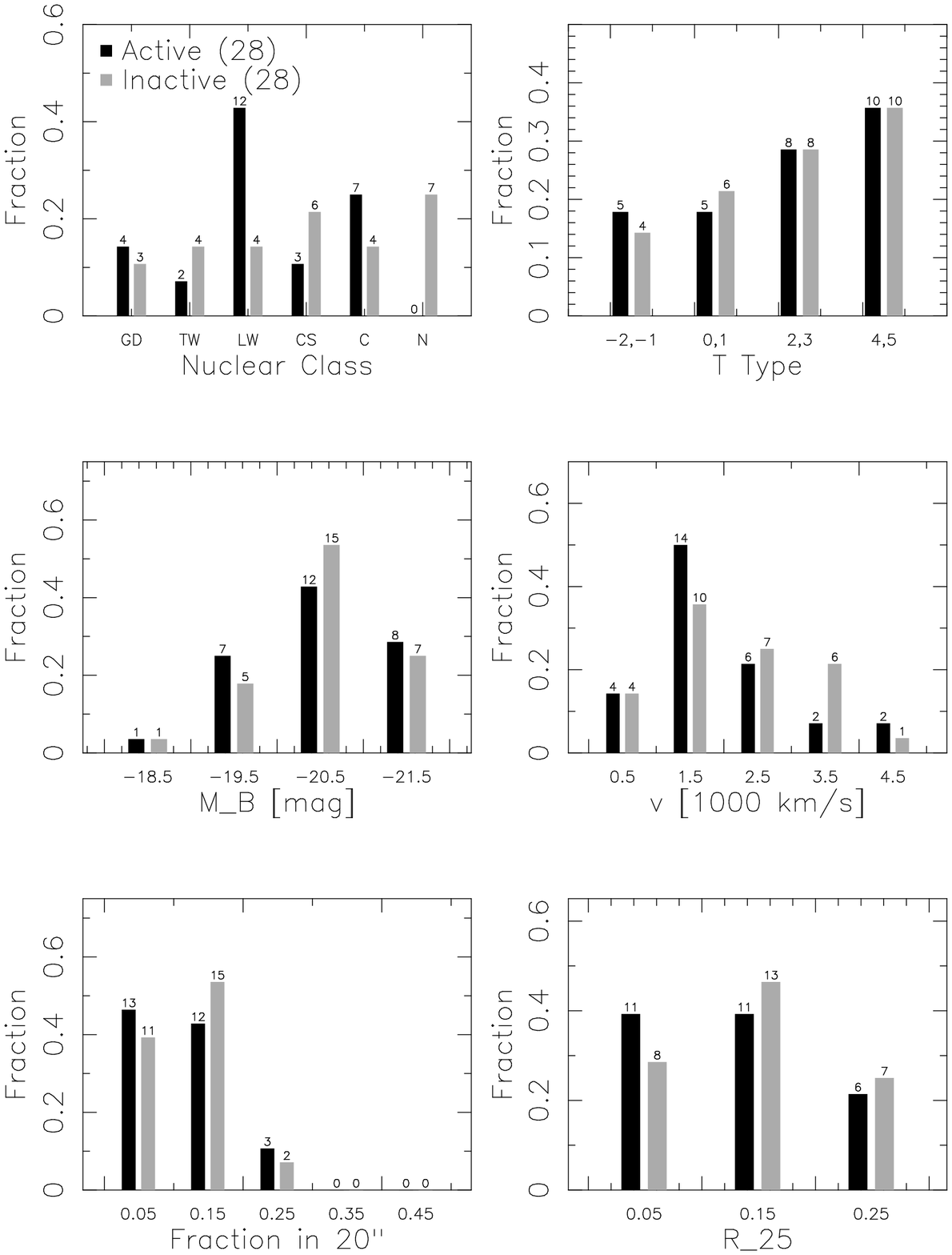}
\caption{
Distribution of the matched sample of 28 active and 28 inactive galaxies 
listed in Table~\ref{tbl:matchagn} and described in \S\ref{sec:match}.  
The upper left histogram shows the distribution of these galaxies into 
the six nuclear classes defined in Paper~I. 
These classes are: GD: grand design nuclear dust spiral; TW: tightly wound 
nuclear spiral; LW: loosely wound nuclear spiral; CS: chaotic spiral; 
C: chaotic dust structure; N: no dust structure present. 
The histogram bars are normalized 
to the total number of galaxies in each class and the number of galaxies 
in each class is given above the histogram bar. The remaining five histograms 
show the distribution of the Hubble type, $B$ luminosity, heliocentric 
velocity, size, and inclination of the active and inactive galaxies. 
The seven inactive galaxies with no circumnuclear dust structure, as opposed 
to no similar active galaxies, has a formal significance of $3\sigma$. 
There are also three times as many active galaxies with loosely wound spirals 
as inactive galaxies, although this is not formally significant. 
The remaining panels show the close correspondence of the 
active and inactive galaxies in each of these five host galaxy properties. 
\label{fig:matchagn} }
\end{figure}

This method successfully found a unique active galaxy match for each 
inactive galaxy with $v < 5000 \kms$, 
$R_{25} < 0.30$, and at least one possible pair: the best possible outcome. 
For other parameter constraints described below it was either equally 
successful or at worst would fail to match only one or two control galaxies. 
This process yielded a final, matched sample of 28 active galaxies and 
28 inactive galaxies. The distribution of nuclear classifications 
and host galaxy properties for this sample is shown in 
Figure~\ref{fig:matchagn}. The galaxies in this active and inactive 
sample are listed in Table~\ref{tbl:matchagn} and their host galaxy 
properties are provided in Paper~I. 
The mean (and rms) differences between the pairs for each of these properties 
are: $\Delta M_B = -0.03 (0.29)$ mag, $\Delta T = 0.11 (0.72)$, 
$\Delta v_h = -218 (553) \kms$, $R_{25} = -0.01 (0.04)$, and 
$frac_{20} = -0.01 (0.04)$. 
If we increased the size of the constraints, 
such as from 0.5 to 1 mag in $M_B$, the sample size would only increase 
by three pairs, which we decided did not justify the chance this would 
increase systematic biases. 
The top left panel of the figure shows the relative distribution of 
active and inactive galaxies in each of the six nuclear classes. 
The remaining panels show the distribution of host galaxy properties. 

We did not include the presence or absence of a large-scale bar in this 
sample selection, as this additional constraint would have further reduced the 
sample size. The sample listed in Table~\ref{tbl:matchagn} has more unbarred
inactive galaxies (ten) than unbarred active galaxies (four). As we show 
next, nuclear spiral structure appears to be present in barred and 
unbarred galaxies with equal frequency, so this should not affect our 
results, although the distribution of barred and unbarred galaxies into the 
nuclear spiral classes is different.  

\begin{figure}
\epsscale{0.85}
\plotone{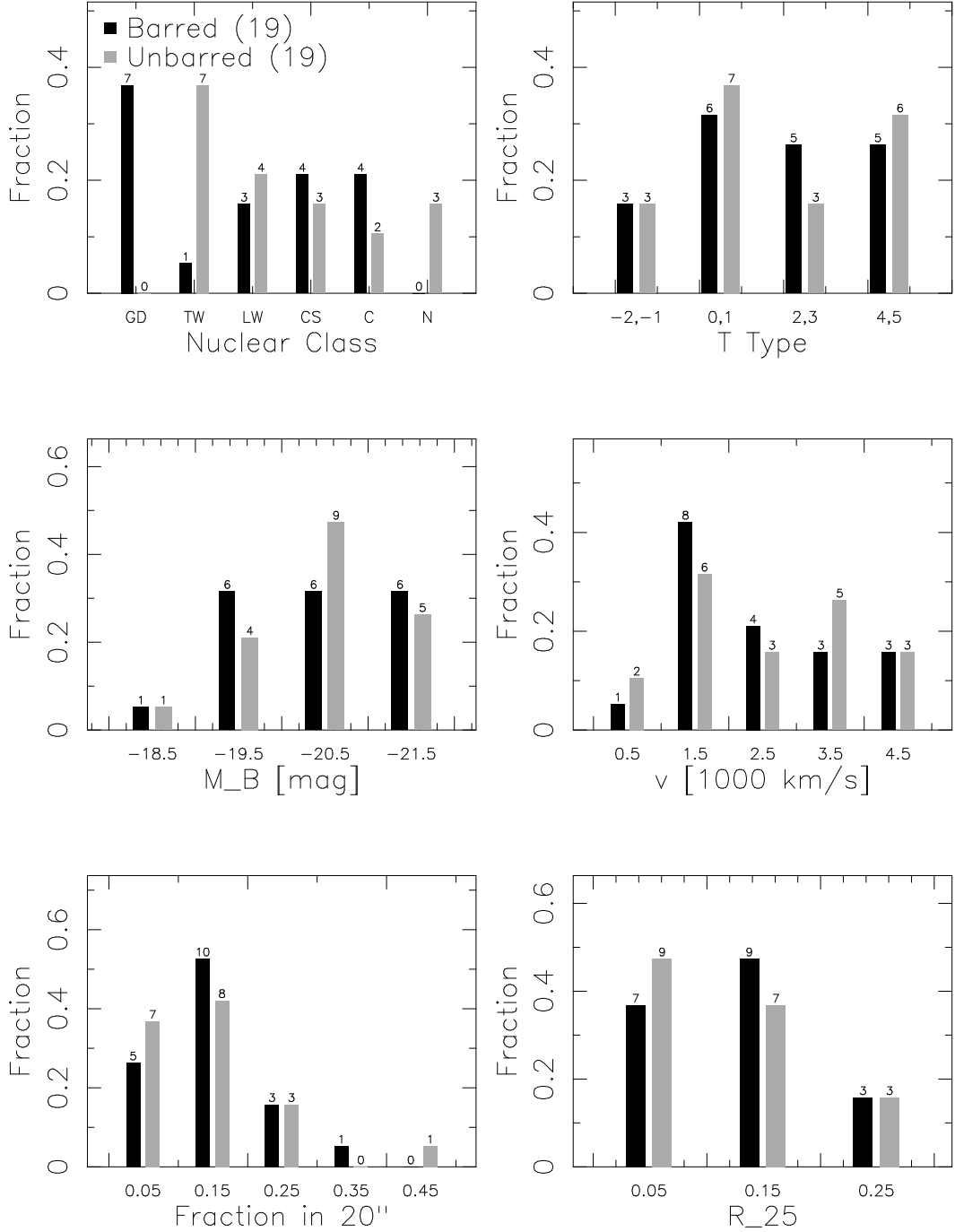}
\caption{Same as Figure~\ref{fig:matchagn} for the matched sample of 19 
barred and 19 unbarred galaxies listed in Table~\ref{tbl:matchbar} and 
described in \S\ref{sec:match}. Grand design nuclear spirals are only found 
in barred galaxies, while there is a marginally significant tendency for 
tightly wound nuclear spirals to be present in unbarred galaxies. The 
remaining panels show the similar host galaxy properties of the barred and 
unbarred galaxies. 
\label{fig:matchbar} }
\end{figure}

We created matched samples of barred and unbarred galaxies in the same manner 
as for the active and inactive galaxies. The goal of this investigation 
was to measure the frequency of nuclear spiral structure in barred and unbarred 
galaxies, as well as to verify the two results noted in Paper~I: 
the existence of grand design nuclear spirals only in barred galaxies and the 
strong tendency for tightly wound nuclear spirals to avoid barred galaxies.  
Due to the small number of unbarred galaxies (25 or 86) in 
this sample, only 19 pairs of barred and unbarred galaxies were identified. 
These galaxies are listed in Table~\ref{tbl:matchbar}. 
Figure~\ref{fig:matchbar} shows the distribution of these galaxies in each 
of the morphological classes for circumnuclear dust, along with the
distributions of their host galaxy properties. The mean differences 
between the pairs of barred and unbarred galaxies in each of the parameters 
are: $\Delta M_B = -0.01 (0.27)$ mag, $\Delta T = 0.19 (0.59)$, 
$\Delta v_h = -68 (528) \kms$, $R_{25} = 0.01 (0.05)$, and 
$frac_{20} = 0.00 (0.04)$. The barred and unbarred pairs were not 
required to have the same nuclear activity, although as eleven of 19 
barred galaxies and eight of 19 unbarred galaxies were active, the 
comparison below should not be biased by any differences between active and 
inactive galaxies. 

\section{Results for Active Galaxies} \label{sec:active}

 \subsection{Frequency of nuclear spiral structure in active galaxies}

The main result of Figure~\ref{fig:matchagn} is that nuclear spiral structure 
is found with comparable frequency in active and inactive galaxies 
for a sample well-matched by host galaxy properties. 
When all four nuclear spiral classes are summed together, 21 of the 28 active 
galaxies (75\%) have nuclear dust spirals compared to 17 of the 28 (61\%) 
inactive galaxies. 
The significance of this result can be estimated with a binomial distribution 
as each galaxiy either has a nuclear spiral or does not, these outcomes are 
statistically independent, and each galaxy has the same probability of 
having a nuclear spiral. 
The variances of the active and inactive samples are 
$\sigma_a^2 = 28 \times 0.75 \times 0.25 = 5.25$ and 
$\sigma_i^2 = 28 \times 0.61 \times 0.39 = 6.66$, respectively, and 
the variance of the difference is $\sigma^2 = \sigma_a^2 + \sigma_i^2 = 11.91$. 
Thus, the observed spiral frequencies formally differ by only $1.2\sigma$, 
which is not statistically significant. 

Of the four types of nuclear spirals, there are approximately equal numbers of 
active and inactive galaxies in the grand design, tightly wound, and chaotic 
spiral classes. The active galaxies with loosely wound nuclear spirals 
outnumber inactive galaxies with similar circumnuclear morphologies 
by a factor of three to one, although this excess has less than 
$2\sigma$ significance. 
The main result from this figure is that nuclear spirals are found with 
comparable frequency in active and inactive galaxies. 

 \subsection{Frequency of circumnuclear dust structures in active galaxies}

Another interesting result from Figure~\ref{fig:matchagn} is that 
25\% (seven) of the inactive galaxies have no obvious circumnuclear dust 
structure at all, whereas all of the active galaxies have some structure in 
their circumnuclear dust distribution. These galaxies must either have no 
circumnuclear dust, or this dust must be smoothly distributed. 
Employment of the binomial distribution estimate shows that this difference 
has a formal significance of $3\sigma$. 
These galaxies include four S0s, an Sa, an Sab, and an Sbc, and therefore it 
is not Hubble type alone which drives the absence of circumnuclear dust 
structure. Spatial resolution is important to identify dust structures 
in the centers of galaxies, as color maps are most sensitive to dust 
structure with an angular size comparable to the PSF. We are therefore 
progressively less sensitive to small-scale dust structure in the more distant 
galaxies. However, as we have an equal number of active and inactive 
galaxies at each distance, there is no bias against the detection of dust 
structures in one sample relative to the other. Therefore spatial resolution 
will not effect the relative frequency of dust structure in active and 
inactive galaxies, although 
it may effect the number of galaxies with detected dust structure, 

The full sample from Paper~I includes these same seven inactive galaxies with 
no circumnuclear dust structure: NGC\,357, NGC\,628, NGC\,1398, NGC\,1638, 
NGC\,3300, NGC\,3458, and NGC\,7096, along with one active galaxy: 
NGC\,3362. This AGN was excluded from the discussion in the present paper 
because it has a heliocentric velocity greater than $5000 \kms$ 
($v \sim 8300$). 
The lack of circumnuclear dust structure in this AGN may be due to poor 
spatial resolution, or contamination by light from the central point source. 
The presence of some emission line gas in the central region of NGC\,3362 
indicates that it does have some circumnuclear gas, even though no dust 
structure is apparent in the colormap. 
The fact that the only AGN in this class in the full sample is at a large 
distance, and in any case shows emission line structure in the circumnuclear 
region, further supports our result that inactive galaxies have a strong 
tendency to lack circumnuclear dust structure relative to active galaxies. 

\section{Results for all Galaxies} \label{sec:all}

 \subsection{Grand-design nuclear spirals in barred galaxies}

The upper left panel of Figure~\ref{fig:matchbar} shows the distribution of 
the barred and unbarred galaxies into each nuclear class. 
While the sample is smaller than shown in Figure~\ref{fig:matchagn}, the 
differences appear more striking. 
The first column confirms the result from Paper~I that grand design nuclear 
spiral structure is only found in galaxies with large-scale bars.  
Seven grand design nuclear spirals are found in barred galaxies, whereas 
none are found in unbarred galaxies. 
This result has $3\sigma$ formal significance. 
In Paper~I we found 19 barred galaxies with grand design structure and no 
unbarred galaxies in this class. 
Grand design nuclear spiral structure is therefore only found in galaxies with 
large-scale bars. 
Furthermore, these grand design nuclear dust spirals always appear to connect 
with dust lanes along the leading edges of the large-scale bar. 
If the barred galaxies in this sample and in the full sample discussed in 
Paper~I are representative of the population of barred galaxies, grand 
design nuclear spirals are present in approximately a third of all barred 
galaxies (27\% of the full sample from Paper~I, 37\% of the present, smaller 
sample). These grand design nuclear spirals are therefore more common than 
inner rings, which were only found in a small minority of barred galaxies in 
Paper~I. Only three of the 19 barred galaxies listed in Table~2 show evidence 
for inner rings in the \hst observations presented in Paper~I. 

The morphology of these dust lanes is in good agreement with the predictions 
of hydrodynamic simulations of gas morphology in the circumnuclear regions of 
barred galaxies \citep{englmaier00,maciejewski02,wada02}. 
The observations clearly show the shock fronts in large-scale bars continues 
into the circumnuclear region to the limit of our angular resolution, which 
corresponds to tens of parsecs from the nucleus. 
The inflow of matter due to large-scale bars can lead to the secular evolution 
of galaxies, 
fuel circumnuclear star formation, accretion, and may eventually 
lead to the dissolution of the large-scale bar \citep{kormendy93,friedli93}. 
The fact that inner rings are only found in a small minority of barred 
galaxies, and are less common than grand design spirals, shows that the 
``pile up'' of bar-driven inflow into a ring is a rare occurence.  

 \subsection{Tightly wound nuclear spirals in unbarred galaxies}

In the full sample described in Paper~I there were three barred galaxies 
(out of 70) with tightly wound nuclear dust spirals compared with eight 
(of 31) unbarred galaxies. 
Two of those barred galaxies with tightly wound nuclear dust spirals were 
excluded from this analysis because of distance or because there 
was not an unbarred galaxy with comparable host galaxy properties. 
The matched sample shown in Figure~\ref{fig:matchbar} has seven unbarred 
galaxies with tightly wound nuclear spirals and only one barred galaxy with a 
tightly wound nuclear spiral. This difference has a formal significance of 
$2.5\sigma$ and therefore may represent a second real difference in the 
circumnuclear structure of barred galaxies. 
As there are a small number of barred galaxies with tightly wound nuclear dust 
spirals, the connection between tightly wound nuclear spirals and the absence 
of a large-scale bar is not as strong as the connection between grand design 
nuclear spirals and large-scale bars. 
Nevertheless, given the fact that barred galaxies far outnumber unbarred 
galaxies in the full sample, and there is an even stronger trend in the 
full sample, there may be a physical connection between the absence of a 
large-scale bar and the presence of a tightly wound nuclear dust spiral. 
The tendency for tightly wound nuclear spirals to be present in unbarred 
galaxies could be confirmed (improved to $3\sigma$ significance) with 
an additional sample of on order ten unbarred galaxies that are well-matched 
to our existing barred galaxy sample. 

 \subsection{Correlations with host galaxy properties}

To investigate any additional correlations between host galaxy properties 
and circumnuclear dust structure, we have examined the mean host galaxy 
properties of all of the galaxies in Paper~I with $R_{25} < 0.30$ and 
$v < 5000 \kms$. 
The mean Hubble type of this sample is only slightly later than Sab. 
This is also the case for all of the individual nuclear classes except for 
the loosely wound spiral class, which has an earlier mean Hubble 
type between S0/a and Sa. 
The similar mean Hubble types for each nuclear class suggests that there is 
not a strong relationship between galaxy type and the morphology of the 
circumnuclear dust. 
The fact that the loosely wound spirals are more common in galaxies whose 
large-scale spiral arms are more tightly wound may be due to the relative 
importance of the bulge and disk in the circumnuclear and large-scale 
morphology. However, \citet{rubin85} showed that the rotation curves of 
galaxies are more dependent on luminosity than on Hubble type. 

A difference in luminosity, parametrized here as $M_B$, could therefore explain 
the variations in circumnuclear dust structure if the amount of differential 
rotation is the primary factor in the pitch angle of the nuclear spirals. 
In general, there are no large differences in the average $M_B$ from class to 
class. The mean $M_B$ for the entire sample is -20.3 mag and the mean value 
for each class deviates at most by little over half a magnitude from this 
value. 
The deviation is largest between the tightly wound nuclear spirals, which have 
$M_B = -20.9$ mag and the loosely wound nuclear spirals, which have 
$M_B = -20.1$ mag. Therefore it appears that the tightly wound nuclear spirals 
are found in more luminous galaxies. As \citet{rubin85} found that 
more luminous galaxies have larger central velocity gradients, 
tightly wound nuclear spirals may therefore be found in more luminous galaxies 
if these galaxies also deviate more strongly from solid-body rotation. 
An alternate explanation is that the tightly wound spirals have been 
undisturbed for a longer period of time and have had more time to wind. 
Kinematic observations are necessary to determine if the circumnuclear 
rotation curve is the primary driver of nuclear spiral pitch angle. 

Two additional parameters that describe the morphology of these galaxies, 
but are not included in our paired sample matching, are the presence of a ring 
or pseudo ring and whether or not a galaxy is classified as peculiar. 
We have used the RC3 classifications listed in Table~3 of Paper~I to compute 
the relative frequency of ringlike and peculiar morphology as a function 
of nuclear classification and found that the relative frequencies of these 
classifications are consistent with each other across the six circumnuclear 
dust classes. This consistency is not very significant, however, 
as the quality of the RC3 classifications is difficult to quantify 
for the full sample. A larger number of well-matched galaxy pairs would 
provide an improved basis for investigation of correlations between 
circumnuclear morphology and host galaxy properties. 

All of the mechanisms proposed to fuel nuclear activity in galaxies are 
also methods that can trigger circumnuclear star formation and may lead to 
secular evolution. Recent 
work on a direct connection between Seyferts and starbursts by 
\citet{storchi01} has investigated connections between the circumnuclear dust 
morphology and star formation. These authors studied the 
relationship between the \citet{malkan98} ``inner Hubble type'' classification 
and the presence of circumnuclear star formation and found that most 
Seyferts with circumnuclear star formation were classified as late-type 
inner spirals, or loosely wound spirals in our classification system. 
If there is more star formation in galaxies with high pitch angle nuclear 
spirals, this could help to explain the slightly brighter $M_B$ of the 
loosely wound nuclear spirals relative to the other three nuclear spiral 
classes. Inspection of the images and colormaps shown in Figure~1 of 
Paper~I shows some observational support for the results of 
\citet{storchi01}, namely the star formation in NGC\,1672, NGC\,4314, 
NGC\,5427, and NGC\,6951. We note, however, that our sample was not designed 
to carry out a well-defined study of star formation. 

\section{Discussion} \label{sec:discuss}

Although our analysis based on matched pairs of either active/inactive or 
barred/unbarred  
galaxies minimizes sample biases, 
there are nevertheless several to consider. 
The first is that the bar and active galaxy classifications are 
nonuniform. As described in Paper~I, several of the allegedly inactive 
galaxies designated as controls for the RSA Seyfert Sample were later 
reclassified as active galaxies by the Palomar Survey \citep{ho97}. 
While many of the inactive galaxies in this sample were part of the Palomar 
survey, the other galaxies were observed as part of less sensitive surveys 
that may be biased against the detection of very low-luminosity AGN. 
This reflects a problem inherent in the 
view of galaxies as either active or inactive, when in reality the 
intensity of nuclear activity varies along a continuum over many orders of 
magnitude. Proper inclusion of, for example, bins of $L_{bol}/L_{Edd}$ or 
accretion rate, would require considerably larger samples. 
For the current investigation it is very nevertheless unlikely that at least 
any bright Seyfert~1 galaxies were included in our inactive sample due to the 
fact that there are no bright, unresolved point sources at the centers of 
their visible or near-infrared \hst images. 
Several quite faint active galaxies from the 
Palomar Survey do not have distinct nuclei, so the possibility remains 
that there are very low-luminosity active galaxies in the inactive sample. 
The vast majority of these misclassified active galaxies 
would have to be in the dust-free or chaotic classes to significantly 
change our results on active galaxies.  

A second potential selection effect is that the brightness of the nuclear 
point source in the active galaxies may increase the $M_B$ used to match the 
active and inactive samples. 
While this is a concern for comparisons of luminous AGN, it is unlikely to 
affect this study because most of the bona fide AGN are still relatively 
low luminosity nuclei and only make a minor contribution to the total 
galaxy$+$AGN luminosity. 
To test this we have artificially decreased the $M_B$ value by half a magnitude 
and rematched the pairs of active and control samples. 
There was no significant difference in either the frequency of 
nuclear spiral structure or the result that the total absence 
of circumnuclear dust was only found in the inactive galaxies. 

A related uncertainty to the first point above is that the sample matching 
process is only as reliable as the host galaxy properties that are used to 
create the samples. 
Fortunately, all of the host galaxy properties were obtained from the 
RC3 catalog \citep{devaucouleurs91} 
and therefore were extracted from a homogeneous dataset. 
While it is more difficult to obtain accurate galaxy classifications for 
more distant or less luminous galaxies, the fact that the galaxy samples 
were matched based on both of these properties simultaneously should 
account for distance- and brightness-dependent errors in the measurement of 
the host galaxy properties. 
A more serious problem is the bar classifications. 
While all of these galaxies have bar classifications 
in the RC3 catalog, it is now well known that NIR observations often identify 
bars in galaxies considered unbarred at visible wavelengths 
\citep[\eg][]{mulchaey97a} and several of the unbarred galaxies lack NIR data. 
If some of the unbarred galaxies were later found to be barred based on 
NIR observations, that would not change our result that grand design nuclear 
spirals are only found in galaxies with large-scale bars, 
although this could effect the result that small pitch angle nuclear 
spirals (the TW class) are preferentially found in unbarred galaxies. 

A final uncertainty is the quality of our own nuclear classifications. 
As discussed in Paper~I, the classifications for many galaxies were initially 
disputed. However, most of these disputes were between the different nuclear 
spiral classes and would not affect our result on the relative frequency of 
nuclear spirals in active and inactive galaxies. The main uncertainty that 
could affect this result is the misclassification between the chaotic spiral 
and chaotic dust classes. However, half (three) of the active galaxies 
classified as chaotic would have to have spiral structure and 
half (also three) of the inactive galaxies classified as chaotic spirals 
would have to lack spiral structure in order produce a $3\sigma$ difference 
between active and inactive galaxies. The probability of both circumstances 
is quite low. 

 \subsection{Matter inflow in barred galaxies}

Approximately a third of the barred galaxies in this sample show curved dust 
lanes which start at the leading edges of their large-scale bars, have a grand 
design morphology in the circumnuclear region 
and then extend into the unresolved nuclear region. 
The apparent extension of the shock fronts in these galaxies into the 
circumnuclear region presents a strong case for the bar-driven inflow of 
matter to the nucleus in at least some barred galaxies. 
While single color maps do not provide a direct measure of the matter surface 
density in these dust lanes due to uncertainties caused by the unknown dust 
geometry and optical depth effects, a lower limit of a few $M_\odot$ pc$^{-2}$ 
is reasonable \citep{regan99,martini99}. 
In simulations of dust lanes in barred galaxies the maximal inflow rates 
are as high as $\sim 100 \kms$ \citep{athanassoula92}, similar to measurements 
of the maximum gas velocity parallel to the dust lanes for several galaxies 
\citep{regan97,reynaud98}. 
The mean inflow in the simulations is a few $\kms$. 

\citet{sakamoto99} derive a lower limit of $0.1 - 1$ M$_\odot$ yr$^{-1}$ 
for the accretion rate into the central kiloparsec of barred galaxies from 
their CO data. 
This amount of matter inflow from bars is sufficient to explain the enhanced 
molecular gas content at the centers of barred galaxies 
\citep{sakamoto99,regan99b} and trigger circumnuclear star formation. 
There may also be sufficient matter inflow to fuel low-luminosity nuclear 
activity. 
However, this material must still travel from tens of parsecs, the best 
spatial resolution of these observations, to scales of less than a parsec in 
order to come within range of the central black hole's gravitational potential. 

 \subsection{Do nuclear dust spirals fuel nuclear activity?}

The presence of grand design spiral shocks demonstrates the presence of a 
mechanism for matter inflow to
the centers of barred galaxies, yet this symmetric spiral structure is 
only found in a minority of active and inactive galaxies. The majority of 
nuclear dust spirals take less symmetric forms with a range in pitch angle, or 
even have a chaotic 
appearance that may just exhibit some shear due to differential rotation. 
These forms of nuclear spiral structure are probably not formed by 
large-scale forces in the host galaxy, such as a bar, and 
instead are probably created {\it in situ}. 
Turbulence, either due to pressure or gravitational forces, may create 
structures that are then sheared into a spiral form due to differential 
rotation \citep{elmegreen98,montenegro99,wada01b}. 
The characteristics of density contrasts formed by acoustic turbulence include 
low density enhancements over the ambient ISM, a lack of associated 
star formation due to their low density, an increase in contrast at smaller 
radii, and that they should completely fill the circumnuclear disks in which 
they propagate \citep{elmegreen02}. 

As the turbulence which leads to these spirals produces shocks in the ISM, 
and shocks dissipate energy and angular momentum, these spirals trace 
the sites of angular momentum transfer that can lead to the fueling 
of nuclear activity. In a test of this scenario by \citet{elmegreen02},  
analysis of nuclear dust spirals in the LINERs NGC\,4450 and NGC\,4736 
show that the azimuthal Fourier power spectrum of these spirals has a slope 
of $-5/3$, indicative of turbulence. The tightly wound, loosely wound, 
and chaotic nuclear spirals have similar morphologies to the nuclear 
spirals studied by \citet{elmegreen98} and \citet{elmegreen02}, which 
suggests that most nuclear spirals are caused by a turbulent process. 

While morphological arguments suggests that nuclear spirals are formed 
by turbulence and shocks, the question remains: Does sufficient angular 
momentum transfer occur to fuel nuclear accretion? The presence of an 
insignificant excess of nuclear spirals in active galaxies over inactive 
galaxies does not support this scenario. 
This is not to say that matter inflow is ruled out, only that it stops short 
of the central black hole. 
Kinematic evidence for inflow associated with these spirals, as well
as emission-line signatures of shocks, would provide more compelling evidence 
that nuclear spiral structure can produce the accretion rates required to 
fuel low-luminosity AGN. 

 \subsection{Constraints on the duty cycle of nuclear activity}

Bars and interactions are mechanisms that can remove angular momentum and 
fuel nuclear activity. Nuclear spirals could be another possibility, provided 
that they trace sufficiently strong shocks in the circumnuclear ISM. 
However, all three of these proposed fueling mechanisms are found with 
approximately equal frequency in both active and inactive galaxies. 
This naturally prompts the question: If these mechanisms are responsible for 
fueling AGN, why are AGN not found in the otherwise indistinguishable 
inactive galaxies? These observations of nuclear dust spirals, as well as 
previous work on other similarities in the host galaxies of low-luminosity 
AGN on larger scales, show that there are not clear differences between 
active and inactive galaxies on any scales greater than approximately a 
hundred parsecs, the smallest scales probed by the present work. 

The absence of clear differences between active and inactive 
galaxies to within approximately 100 pc radius of their galactic nuclei, 
combined with the presumption that all of these galaxies possess 
supermassive black holes, suggests that the characteristic inflow timescale 
from this radius is an upper limit on the lifetime of nuclear activity. 
A plausible estimate of this inflow timescale is the dynamical timescale. 
At a hundred parsecs, the dynamical timescale is several million years in a 
typical galaxy. If the structures responsible for AGN fueling exist at 
hundreds of parsec scales, but they are equally common in inactive galaxies, 
this suggests that the lifetime for an individual episode of nuclear activity 
is less than several million years. In our proposed scenario, where the 
structures responsible for AGN fueling are present in essentially all 
galaxies, all galaxies periodically go through a short-lived AGN phase. 

\section{Conclusions} \label{sec:conc}

The cold ISM in the circumnuclear region of nearly all spiral galaxies 
exhibits a wealth of structure. 
This structure often takes the form of nuclear dust spirals, which 
demonstrates that some differential rotation is present in the circumnuclear 
region of most spiral galaxies. 

A comparison of a well-matched sample of 19 barred and 19 unbarred galaxies has 
confirmed our previous result (Paper~I) that grand design nuclear dust 
spirals are only found in galaxies 
with large-scale bars. Grand design spirals are present in approximately 
a third of all barred galaxies and these circumnuclear spirals 
connect to the relatively straight dust lanes associated with the large-scale 
bar. The dust lanes therefore form essentially contiguous structures from 
many kiloparsec scales to within tens of parsecs of galactic centers. 
These nuclear spirals are also more common than inner rings, which are 
found in only a minority of all barred galaxies. The fact that most barred 
galaxies do not have rings inside the bar suggests that bar-driven 
inflow does not commonly stall in a circumnuclear ring. 
Our comparison of barred and unbarred galaxies also shows that there is a 
marginally significant tendency for tightly wound nuclear dust spirals to be 
found in galaxies that lack large-scale bars. The low pitch angle of these 
spirals may be due to an increased time for winding or greater differential 
rotation.

To test for differences in the circumnuclear dust distribution in active 
and inactive galaxies, we have created a well-matched sample of 28 active 
and 28 inactive galaxies by identifying pairs with all of the same host 
galaxy properties. 
This analysis finds only a statistically insignificant excess of nuclear 
spirals in active galaxies over the inactive control sample and that 
nuclear spiral structure is present in the majority of both types. 
A perhaps more surprising result was that although all of the active galaxies 
have some discernible dust structure, seven of the 28 inactive galaxies have 
no circumnuclear dust structure.  These galaxies either have an extremely 
smooth dust distribution or no cold component to their ISM in the 
circumnuclear region. 
The comparable frequency of nuclear dust spirals in a well-matched sample of 
active and inactive galaxies, when combined with similar results for the 
frequency of bars or interactions from previous work, suggests that there is 
no universal fueling mechanism for low-luminosity AGN at scales greater than 
100 pc from galactic nuclei.

We have shown that the circumnuclear dust morphologies of most active 
and inactive galaxies are indistinguishable.  
All of these galaxies are assumed to contain supermassive black holes 
and therefore the main difference between them is whether or not that 
central black hole is currently fueled. 
If the structures responsible for fueling nuclear activity are not 
present on the hundreds of parsec scales probed by these observations, or 
are present with equal frequency in active and inactive galaxies, then 
the timescale for nuclear activity must be less than the timescale for 
matter inflow from these spatial scales. 
If the dynamical timescale on these spatial scales is an approximate measure of 
the characteristic inflow time, this suggests an upper limit to the lifetime 
of an individual episode of nuclear activity is several million years. 

\acknowledgements 

Support for this work was provided by NASA through grant numbers 
GO-7330, GO-7867, and GO-08597 from the Space Telescope Science Institute, 
which is operated by the Association of Universities for Research in Astronomy, 
Inc., under NASA contract NAS5-26555.  
PM was supported by a Carnegie Starr Fellowship. 
We would like to thank Andy Gould, Witold Maciejewski, Keiichi Wada, and the 
referee for helpful suggestions and comments. 
This research has made use of the NASA/IPAC Extragalactic Database (NED) which 
is operated by the Jet Propulsion Laboratory, California Institute of 
Technology, under contract with the National Aeronautics and Space 
Administration. 

{}

\clearpage

\clearpage

\clearpage

\begin{center}
\begin{deluxetable}{cc}
\tablecolumns{2}
\tablenum{1}
\tablewidth{5.0truein}
\tablecaption{Matched active and inactive sample\label{tbl:matchagn}}
\tablehead{
\colhead{Active Galaxy} &
\colhead{Inactive Galaxy} 
}
\startdata
IC2560  &  NGC7392  \\
MRK1066  &  NGC3300 \\ 
NGC1068  &  NGC1398  \\
NGC1667  &  NGC0214  \\
NGC1672  &  NGC5383  \\
NGC2273  &  NGC5691  \\
NGC2336  &  NGC1530  \\
NGC3227  &  IC5267  \\
NGC3486  &  NGC1300  \\
NGC3516  &  NGC1638  \\
NGC3786  &  NGC2179  \\
NGC4143  &  NGC3458  \\
NGC4151  &  NGC2146  \\
NGC4303  &  NGC4254  \\
NGC4725  &  NGC2985  \\
NGC4939  &  NGC3145  \\
NGC5273  &  NGC3032  \\
NGC5347  &  NGC7096  \\
NGC5427  &  NGC2276  \\
NGC5643  &  NGC4030  \\
NGC5953  &  NGC2460  \\
NGC6300  &  NGC3368  \\
NGC6744  &  NGC5054  \\
NGC6814  &  NGC0628  \\
NGC6951  &  NGC5970  \\
NGC7469  &  NGC5614  \\
NGC7496  &  NGC7716  \\
NGC7743  &  NGC0357  \\
\enddata
\tablecomments{
The sample of active galaxies (Col.~1) and the corresponding inactive, 
control for each (Col.~2). Each control galaxy has approximately 
the same host galaxy type, absolute blue luminosity, heliocentric velocity, 
size, and inclination as the active galaxy in the same row of the Table
(see \S\ref{sec:match}). 
The distribution of their nuclear classifications and host galaxy properties 
are shown in Figure~\ref{fig:matchagn}. 
}
\end{deluxetable}
\end{center}

\clearpage

\begin{center}
\begin{deluxetable}{cc}
\tablecolumns{2}
\tablenum{2}
\tablewidth{5.0truein}
\tablecaption{Matched barred and unbarred sample\label{tbl:matchbar}}
\tablehead{
\colhead{Barred Galaxy} &
\colhead{Unbarred Galaxy} 
}
\startdata 
MARK1066  &  IC5063  \\
NGC1530  &  NGC5054  \\
NGC2273  &  IC5267  \\
NGC2276  &  NGC5427  \\ 
NGC2460  &  NGC5953  \\
NGC3032  &  NGC5273  \\
NGC3145  &  NGC4939  \\
NGC3351  &  NGC4380  \\
NGC3516  &  NGC1638  \\
NGC4253  &  MARK1210  \\
NGC4303  &  NGC4254  \\
NGC4725  &  NGC2985  \\
NGC5135  &  NGC5614  \\
NGC5347  &  NGC7096  \\
NGC5643  &  NGC4030  \\
NGC5691  &  NGC2179  \\
NGC6412  &  NGC0628  \\
NGC7130  &  NGC0788  \\
NGC7469  &  NGC5548  \\
\enddata
\tablecomments{
Same as Table~\ref{tbl:matchagn} for the barred and unbarred galaxy sample 
described in the text and shown in Figure~\ref{fig:matchbar}. 
}
\end{deluxetable}
\end{center}

\end{document}